\documentclass[12pt]{article}
\usepackage{graphicx}

\newcommand\pubdate{\today}

\newcommand\pubnumber{change/delete REPORT-\#}

\def\Title#1{\begin{center} {\Large #1 } \end{center}}
\def\Author#1{\begin{center}{ \sc #1} \end{center}}
\def\Address#1{\begin{center}{ \it #1} \end{center}}

\newcommand\pubblock{\rightline{\begin{tabular}{l} \pubnumber\\
         \pubdate  \end{tabular}}}
\newenvironment{Abstract}{\begin{center}{\bf Abstract}\end{center} \bigskip \begin{quotation}  }{\end{quotation}}
\newenvironment{Presented}{\begin{quotation} \begin{center} 
             PRESENTED AT\end{center}\bigskip 
      \begin{center}\begin{large}}{\end{large}\end{center} \end{quotation}}

\def\beq{\begin{equation}}
\def\eeq#1{\label{#1}\end{equation}}
\def\eeqn{\end{equation}}

\def\beqa{\begin{eqnarray}}
\def\eeqa#1{\label{#1}\end{eqnarray}}
\def\eeqan{\end{eqnarray}}

\let\bar=\overbar

\def\Dslash{\not{\hbox{\kern-4pt $D$}}}
\def\dslash{\not{\hbox{\kern-2pt $\del$}}}

\def\msb{{\bar{\ssstyle M \kern -1pt S}}}

\begin{document}
\begin{titlepage}
\pubblock

\vfill

%%%%%%%%%%%%%%%%%%%%%%%%%%%%%%%%%%%%%%%%%%%%%%%%%%%%%%%
%%MODIFY
%%%%%%% TITLE, AUTHOR, ADDRESS 
%%%%%%%%%%%%%%%%%%%%%%%%%%%%%%%%%%%%%%%%%%%%%%%%%%%%%%%

\Title{Measurements of the CKM angle $\gamma$/$\phi_{3}$ at $B$-factories}
\vfill
\Author{Yoshiyuki Onuki on behalf of the Belle and BaBar collaborations}  
\Address{Department of Physics, Tohoku University, 6-3 Aramaki-Aza, Aoba, Sendai, Miyagi, 980-8578, Japan}
\vfill

%%%%%%%%%%%%%%%%%%%%%%%%%%%%%%%%%%%%%%%%%%%%%%%%%%%%%%%
%%MODIFY
%%%%%%% Abstract
%%%%%%%%%%%%%%%%%%%%%%%%%%%%%%%%%%%%%%%%%%%%%%%%%%%%%%%

\begin{Abstract}

The CKM angle $\gamma$/$\phi_{3}$ have been measured by two B-factories,
the PEPII collider for the BaBar experiment and the KEKB collider for the
Belle experiments. 
The present paper reports recent progress in $\gamma$/$\phi_{3}$.

\end{Abstract}

\vfill

\begin{Presented}
The Ninth International Conference on\\
Flavor Physics and CP Violation\\
(FPCP 2011)\\
Maale Hachamisha, Israel,  May 23--27, 2011
\end{Presented}
\vfill

\end{titlepage}
\def\thefootnote{\fnsymbol{footnote}}
\setcounter{footnote}{0}
%

%%%%%%%%%%%%%%%%%%%%%%%%%%%%%%%%%%%%%%%%%%%%%%%%%%%%%%%
%%%%%%% Article body
%%%%%%%%%%%%%%%%%%%%%%%%%%%%%%%%%%%%%%%%%%%%%%%%%%%%%%%

\section{Introduction}
The $CP$ violation in the Standard Model
is explained by the quark mixing matrix of 
the charged current weak interaction. 
The matrix is as known as Cabibbo-Kobayashi-Maskawa(CKM) 
matrix~\cite{ref:ckm}. 
The element has the irreducible complex phase which causes $CP$ violation
in $K$ and $B$ meson 
systems~\cite{ref:K, ref:BaBar_sin2phi1, ref:Belle_sin2phi1}. 
The unitarity of CKM matrix has bring some constrains of the elements 
$V_{i,j}$ to draw triangles in the complex plane,
where $i$ and $j$ stand for the up-type quark $(u,c,t)$ and 
down type quark $(d,s,b)$, respectively. One of the triangles which has
the sides in same order of magnitude is suitable place to measure the 
elements precisely. The angles, $\alpha$/$\phi_{2}$, $\beta$/$\phi_{1}$
and $\gamma$/$\phi_{3}$ has been measured by two $B$-factories, 
the PEPII collider for the BaBar experiment~\cite{ref:BaBar_detector} and 
the KEKB collider for the Belle experiments~\cite{ref:Belle_detector},where
$\alpha/\phi_{2} = Arg[-(V_{td}V_{Vtb}^{*})/(V_{ud}V_{ub}^{*})]$,
$\beta/\phi_{1}  = Arg[-(V_{cd}V_{cb}^{*})/(V_{td}V_{tb}^{*})]$ and
$\gamma/\phi_{3} = Arg[-(V_{ud}V_{ub}^{*})/(V_{cd}V_{cb}^{*})]$.

Current most worst precision of the angle is angle $\gamma$/$\phi_{3}$. 
Improvement of the precision is eagerly awaited.
The $\gamma$/$\phi_{3}$ can be obtained mainly using the process 
$B\rightarrow D^{(*)}K^{(*)}$ involved with interference with
Cabibbo-suppressed $b\rightarrow u$ and Cabibbo-favored $b\rightarrow c$ 
quark transition~\cite{ref:sanda}. 
Some methods to extract $\gamma$/$\phi_{3}$ had been suggested so far:
GLW~\cite{ref:GLW}, ADS~\cite{ref:ADS},  Dalitz~\cite{ref:ggsz, ref:bondar}
analyses.
The GLW analysis uses $D$ meson decays to the $CP$ eigenstates.
The ADS analysis uses Cabibbo-favored $\bar{D}^{0}$ and 
doubly Cabibbo-suppressed $D^{0}$ meson decays.
The Dalitz method uses Cabibbo-allowed three-body $D$ meson decays.

\section{GLW analysis and the recent measurements}

Gronau, London and Wyler (GLW)~\cite{ref:GLW} 
proposed analysis method to extract the angle $\gamma$/$\phi_{3}$ 
with $D^{0}$ and $\bar{D}^{0}$ decay into $CP$ 
eigenstate such as $K^{+}K^{-}$ or $K_{S}\pi^{0}$, etc.
The observables, double ratio and asymmetry, are defined as below.
\begin{eqnarray}
\displaystyle
\begin{array}{rll}
R_{CP\pm} &\equiv & 2\displaystyle\frac{{\cal B}(B^{-}\rightarrow D_{CP\pm}K^-)  + {\cal B}(B^{+}\rightarrow D_{CP\pm}K^{+})}{{\cal B}(B^{-}\rightarrow D^0K^{-})  + {\cal B}(B^{+}\rightarrow \bar{D}^0K^{+})}\\
& = & 1 + r_{B}^{2} \pm 2r_{B}\cos\delta_{B}\cos\phi_{3}
\end{array}
\end{eqnarray}
\begin{eqnarray}
\displaystyle
\begin{array}{rll}
A_{CP\pm} &\equiv &\displaystyle\frac{{\cal B}(B^{-}\rightarrow D_{CP\pm}K^-) - {\cal B}(B^{+}\rightarrow D_{CP\pm}K^+)}{{\cal B}(B^{-}\rightarrow D_{CP\pm}K^-) + {\cal B}(B^{+}\rightarrow D_{CP\pm}K^+)}\\
& = & \pm2r_{B}\sin\delta_{B}\sin\phi_{3}/R_{CP\pm}
\end{array}
\end{eqnarray}
where the $D_{CP\pm}$, $r_{B}$ and $\delta_{B}$ represents 
the $D$ meson decay into the $CP$ eigenstate of even(+) and odd(-), 
ratio of amplitudes between $B^{-}\rightarrow \bar{D}^{0}K^{-}$ and
$B^{-}\rightarrow D^{0}K^{-}$ defined
 $r_{B} \equiv |A(B^{-}\rightarrow \bar{D}^{0}K^{-})/|A(B^{-}\rightarrow D^{0}K^{-})|$, 
the difference of strong phase, respectively.

The $B^{\pm}\rightarrow D^{(*)}_{CP\pm}K^{(*)\pm}$ had been measured 
BaBar~\cite{ref:babar_glw_dk,ref:babar_glw_dstark,ref:babar_glw_dkstar} and 
Belle~\cite{ref:belle_glw_dk}
collaboration.
The BaBar recently has reported the measurement of 
$B^{\pm}\rightarrow D_{CP\pm}K^{\pm}$
following $D_{CP+}\rightarrow K^{+}K^{-}$ and $\pi^{+}\pi{-}$ for $CP$-even, 
$D_{CP-}\rightarrow K_{S}\pi^{0}, K_{S}\pi^{0}, K_{S}\omega$ and $K_{S}\phi$
for $CP$-odd eigenstate with the full dataset of
$467\times 10^{6}$ $\Upsilon(4S)\rightarrow B\bar{B}$ 
decays~\cite{ref:babar_glw}. 
The results of combined $D$ sub-decays as for the each $CP$ eigenstate are 
as below.
\begin{eqnarray}
\displaystyle
\begin{array}{rcccccc}
A_{CP+} & = & 0.25  & \pm & 0.06 & \pm & 0.02 \\
A_{CP-} & = & -0.09 & \pm & 0.07 & \pm & 0.02 \\
R_{CP+} & = & 1.18  & \pm & 0.09 & \pm & 0.05 \\
R_{CP-} & = & 1.07  & \pm & 0.08 & \pm & 0.04 
\end{array}
\end{eqnarray}
where the first error is statistical, the second error is systematic.
The parameter $A_{CP+}$ is different from zero with a
significance of $3.6 \sigma$ standard deviations.
The result indicates the direct $CP$ violation followed 
Dalitz analysis results in the latter Section.
Also, they remove the $D_{CP-}\rightarrow K_{S}\phi$ subsample
with those from the $B^{\pm}\rightarrow DK^{\pm}$, 
$D\rightarrow K_{S}h^{+}h^{-}$ to compare the Dalitz 
analysis~\cite{ref:babar_model-dep3}.
The measured $A_{CP\pm}$ and $R_{CP\pm}$ parameters can 
determine the $x_{\pm}$ and $y_{\pm}$ defined in Eq.~\ref{xy}
with the relation as below.
\begin{eqnarray}
\begin{array}{rcc}
x_{\pm} & = & [R_{CP+}(1 \mp A_{CP+}) - R_{CP-}( 1 \mp A_{CP-})]/4\\
\end{array}
\end{eqnarray}
The obtained $x_{\pm}$ are consistent with the Dalitz analysis result as below.
\begin{eqnarray}
\begin{array}{rcccccc}
x_{+} & = & -0.057 & \pm & 0.039 & \pm & 0.015,\\
x_{-} & = &  0.132 & \pm & 0.042 & \pm & 0.018.\\
\end{array}
\end{eqnarray}

\section{ADS analysis and the recent measurements}

Atwood, Dunietz and Soni (ADS)~\cite{ref:ADS} proposed analysis method 
to extract the angle $\gamma$/$\phi_{3}$ with 
Cabibbo-favored $\bar{D}^{0}$ decay($CFD$) and 
doubly Cabibbo-suppressed $D^{0}$ decay($DCSD$) to adjust the interfering
amplitudes have comparable magnitudes through the 
$B^{-}\rightarrow D^{(*)}K^{(*)-}$ decays, where $D^{(*)}$ means 
$D^{(*)0}$ and $\bar{D}^{(*)0}$.
The final state $f$ of $CFD$($DCFD$) from $D$ meson can be used for ADS 
such as $D^{0}\rightarrow K^{-}\pi^{+}$, $K^{-}\pi^{+}\pi^{0}$
($D^{0}\rightarrow K^{+}\pi^{-}$, $K^{+}\pi^{-}\pi^{0}$), etc.
The observables, double ratio and asymmetry, are defined as below.

\begin{eqnarray}
\displaystyle
\begin{array}{rll}
R_{ADS} &\equiv & \displaystyle\frac{{\cal B}(B^{-}\rightarrow [f]_{D}K^-)  + {\cal B}(B^{+}\rightarrow [\bar{f}]_{D}K^{+})}{{\cal B}(B^{-}\rightarrow [\bar{f}]_{D}K^{-})  + {\cal B}(B^{+}\rightarrow [f]_{D}K^{+})}\\
& = & r_{B}^{2} + r_{D}^{2} + 2r_{B}r_{D}\cos(\delta_{B}+\delta_{D})\cos\phi_{3}
\end{array}
\end{eqnarray}

\begin{eqnarray}
\displaystyle
\begin{array}{rll}
A_{ADS} &\equiv & \displaystyle\frac{{\cal B}(B^{-}\rightarrow [f]_{D}K^-)  - {\cal B}(B^{+}\rightarrow [\bar{f}]_{D}K^{+})}{{\cal B}(B^{-}\rightarrow [\bar{f}]_{D}K^{-})  + {\cal B}(B^{+}\rightarrow [f]_{D}K^{+})}\\
& = & 2r_{B}r_{D}\sin(\delta_{B}+\delta_{D})\sin\phi_{3}/R_{ADS}
\end{array}
\end{eqnarray}
where $r_{D}=|A(D^{0}\rightarrow f)/A(\bar{D}^{0}\rightarrow f)|$ and
$\delta_{D}$ is strong phase difference between 
$\bar{D}^{0}\rightarrow f$ and $D^{0}\rightarrow f$.

The $B^{\pm}\rightarrow D^{(*)}K^{(*)\pm}$ had been measured by
BaBar~\cite{ref:babar_ads_dk1,ref:babar_ads_dk2} and 
Belle collaboration~\cite{ref:belle_ads_dk1}.
The BaBar recently has reported the measurement of 
$B^{\pm}\rightarrow D^{(*)}K^{\pm}$ following 
$D^{*}\rightarrow D\gamma, D\pi^{0}$ and $D\rightarrow K\pi$ 
with the full dataset of
$467\times 10^{6}$ $\Upsilon(4S)\rightarrow B\bar{B}$ 
decays~\cite{ref:babar_ads_dk3}. 
The results are as below.

\begin{eqnarray}
\displaystyle
\begin{array}{rcccccc}
A_{DK} & = & -0.86  & \pm & 0.47 & ^{+}_{-} & ^{0.12}_{0.16} \\
R_{DK} & = & 0.011  & \pm & 0.006 & \pm & 0.002 \\
A^{*}_{(D\gamma)K} & = & -0.36  & \pm & 0.94 & ^{+}_{-} & ^{0.25}_{0.41} \\
R^{*}_{(D\gamma)K} & = & 0.013  & \pm & 0.014 & \pm & 0.008 \\
A^{*}_{(D\pi^{0})K} & = & 0.77  & \pm & 0.35 & \pm & 0.12 \\
R^{*}_{(D\pi^{0})K} & = & 0.018  & \pm & 0.009 & \pm & 0.004
\end{array}
\end{eqnarray}

The BaBar also has updated the $B^{\pm}\rightarrow DK^{\pm}$
following $D\rightarrow K\pi\pi^{0}$ with dataset of 
$474 \times 10^{6}$ $\Upsilon(4S)\rightarrow B\bar{B}$ 
decays~\cite{ref:babar_ads_dk_kpipi0}.
The result is 
\begin{eqnarray}
\displaystyle
\begin{array}{rcccccc}
R_{ADS} & = & (9.1  & ^{+}_{-} & ^{8.2}_{7.6} & ^{+}_{-} & ^{1.4}_{3.7})\times 10^{-3} \\
\end{array}
\end{eqnarray}

The Belle also has reported the measurement of $B^{\pm}\rightarrow DK^{\pm}$
following $D\rightarrow K\pi$ with the full dataset of
$772\times 10^{6}$ $\Upsilon(4S)\rightarrow B\bar{B}$ decays~\cite{ref:belle_ads_dk2}. 
The results are as below.
\begin{eqnarray}
\displaystyle
\begin{array}{rcccccc}
A_{DK} & = & -0.39  & ^{+}_{-} & ^{0.26}_{0.28} & ^{+}_{-} & ^{0.04}_{0.03} \\
R_{DK} & = & 0.0163  & ^{+}_{-} & ^{0.0044}_{0.0041} & ^{+}_{-} & ^{0.0007}_{0.0013}\\
\end{array}
\end{eqnarray}
The measured $R_{DK}$ indicates the first evidence of the signal with a 
significance of $4.1 \sigma$ standard deviations.

\section{\label{Dalitz} Dalitz analysis and the recent measurements}

Dalitz analysis with $D$ meson decay into 
$CP$ eigenstate of three-body decay $K_{S}h^{+}h^{-}$ proposed 
by Giri, Grossman, Soffer, Zupan~\cite{ref:ggsz} and 
Bondar~\cite{ref:bondar} as a effective method to extract the angle 
$\gamma$/$\phi_{3}$, where $h^{\pm}$ represents charged light hadrons 
such as pion and kaon.
The advantage of the method is only use of Cabibbo-allowed $D$ decays. 

\subsection{Model-dependent Dalitz analysis and measurements}

The BaBar and Belle had reported $\gamma$/$\phi_{3}$ measurement
with the model-dependent Dalitz 
analysis~\cite{ref:babar_model-dep1,ref:belle_model-dep1,ref:babar_model-dep2}.
The model-dependent Datliz analysis uses the isobar model~\cite{ref:isobar}
which assume three-body decayed $D$ meson proceed through the intermediate 
two-body resonances. 
The total amplitude of Dalitz plane can be represent with two amplitudes of 
$D^{0}$ and $\bar{D}^{0}$ decays into same final state of $K_{S}h^{+}h^{-}$ 
as below.
\begin{eqnarray}
\displaystyle
\begin{array}{r}
f_{B^{+}} = f_{D}(m_{+}^{2}, m_{-}^{2}) + r_{B}e^{i\phi_{3}+i\delta_{B}}f_{D}(m_{-}^{2},m_{+}^{2})\\
\end{array}
\end{eqnarray}
where $m_{+}^{2} = m_{K_{S}h^{+}}^{2}$, $m_{-}^{2} = m_{K_{S}h^{-}}^{2}$.
The $f_{D}(m_{+}^{2},m_{-}^{2})$ consists of the summed amplitudes of
intermediates which comes in the Dalitz plane and single non-resonant 
amplitude as follows.
\begin{eqnarray}
\displaystyle
\begin{array}{r}
f_{D}(m_{+}^{2},m_{-}^{2}) = \displaystyle\sum^{N}_{j=1}a_{j}e^{i\xi_{j}}{\cal A}_{j}(m_{+}^{2},m_{-}^{2})+a_{NR}e^{i\xi_{NR}} 
\end{array}
\end{eqnarray}
Where 
$a_{j}$ and $\xi_{j}$ are the amplitude and phase of the matrix element,
${\cal A}_{j}$ is the matrix element of the $j$-th resonance, and $a_{NR}$ and
$\xi_{NR}$ are the amplitude and phase of the non-resonant component.

The Belle had reported the result of $B^{\pm}\rightarrow D^{(*)}K^{\pm}$, 
$K_{S}\pi^{+}\pi^{-}$
using 18 two-body amplitudes of isobar model with the dataset of 
$657 \times 10^{6}$ $\Upsilon(4S)\rightarrow B\bar{B}$ decays
~\cite{ref:belle_model-dep2}.

\begin{eqnarray}
\displaystyle
\begin{array}{rcccccccc}
\phi_{3} & = & (78.4  & ^{+}_{-} & ^{10.8}_{11.6} & \pm & 3.6 & \pm & 8.9)^\circ\\
\end{array}
\end{eqnarray}
where the first error is statistical, the second is systematic, the third error is model uncertainty.

The BaBar has reported the result of $B^{\pm}\rightarrow D^{(*)}K^{\pm}$, 
$D^{*}\rightarrow D\pi^0$ and $D\gamma$, 
$D\rightarrow K_{S}\pi^{+}\pi^{-}$ and $K_{S}K^{+}K^{-}$
using isobar model with improved 8 two-body amplitudes 
in each $D$ meson sub-decays with the dataset of 
$468 \times 10^{6}$ $\Upsilon(4S)\rightarrow B\bar{B}$ decays
~\cite{ref:babar_model-dep3}.
\begin{eqnarray}
\displaystyle
\begin{array}{rcccccccc}
\gamma & = & (68  & \pm & 14 & \pm & 4 & \pm & 3)^\circ\\
\end{array}
\end{eqnarray}
where the first error is statistical, the second is systematic, the third error is the model uncertainty.

\subsection{Model-independent Dalitz analysis and measurements}

The model-independent Dalitz analysis had proposed by Giri, Grossman,
Soffer and Zupan~\cite{ref:ggsz}, and further developed by Bonder and Poluektov
~\cite{ref:bonder_model-indep1,ref:bonder_model-indep2}.

Assume that the Dalitz plot is divided into $2\cal N$ bins 
symmetrically to the exchange $m_{-}^{2} \leftrightarrow m_{+}^{2}$.
The bins are denoted by the index $i$ ranging from $-\cal N$ to $\cal N$
(excluding 0); the exchange $m_{+}^{2}\leftrightarrow m_{-}^{2}$
corresponds to the exchange $i \leftrightarrow -i$. Then the expected
number of events in the bins of the Dalitz plot of $D$ from 
$B^{+}\rightarrow DK^{+}$ is

\begin{eqnarray}
\displaystyle
\begin{array}{rcc}
N_{i}^{\pm} & = & h_{B}( K_{i} + r_{B}^{2}K_{-i} + 2\sqrt{K_{i}K_{-i}}(x_{\pm}c_{i}+y_{\pm}s_{i}))\\
\end{array}
\end{eqnarray}
where $K_{i}$ is the number of events in the bins in the Dalitz plot of
the $D^{0}$ in a flavor eigenstate, $h_{B}$ is the normalization constant,
$x_{\pm}$ and $y_{\pm}$ are
\begin{eqnarray}
\displaystyle
\begin{array}{rcc}
x_{\pm}=r_{B}\cos(\delta_{B}\pm\phi_{3}),\\ 
y_{\pm}=r_{B}\sin(\delta_{B}\pm\phi_{3}).
\end{array}
\label{xy}
\end{eqnarray}
Coefficients $c_{i}$ and $s_{i}$, which include the information about
the cosine and sine of the phase difference given by
\begin{eqnarray}
\begin{array}{rcc}
c_{i} & = & \frac{ \int_{{\cal D}_{i}} \sqrt{ p_{D} \bar{p}_{D} }\cos(\Delta \delta_{D}(m_{+}^{2},m_{-}^{2})d{\cal D}}{\sqrt{\int_{{\cal D}_{i}} p_{D}d{\cal D} \int_{{\cal D}_{i}}\bar{p}_{D}d{\cal D}}},\\
s_{i} & = & \frac{ \int_{{\cal D}_{i}} \sqrt{ p_{D} \bar{p}_{D} }\sin(\Delta \delta_{D}(m_{+}^{2},m_{-}^{2})d{\cal D}}{\sqrt{\int_{{\cal D}_{i}} p_{D}d{\cal D} \int_{{\cal D}_{i}}\bar{p}_{D}d{\cal D}}}.
\end{array}
\end{eqnarray}
These averaged strong phases, $c_{i}$ and $s_{i}$, in each bin can be 
extracted from the quantum-correlated $D^{0}$ decays from 
$\psi(3770)\rightarrow D\bar{D}$ process. 
The measurement had been performed by CLEO collaboration~\cite{ref:cleo_cisi}.

The first model-independent Dalitz analysis of 
$B^{\pm}\rightarrow DK^{\pm}$, $D\rightarrow K_{S}\pi^{+}\pi^{-}$
has been performed by Belle~\cite{ref:belle_model-indep}.
The results are as below
\begin{eqnarray}
\displaystyle
\begin{array}{rcccccccc}
x_{-} & = & +0.095 & \pm & 0.045 & \pm  & 0.014 & \pm & 0.017, \\
y_{-} & = & +0.137 & ^{+}_{-} & ^{0.0053}_{0.057} & \pm  & 0.019 & \pm & 0.029,\\
x_{+} & = & -0.110 & \pm & 0.043 & \pm  & 0.014 & \pm & 0.016, \\
y_{+} & = & -0.050 & ^{+}_{-} & ^{0.052}_{0.055} & \pm  & 0.011 & \pm & 0.021,
\end{array}
\end{eqnarray}
where the first error is statistical, the second is systematic, the third error is the uncertainty from $c_{i}$ and $s_{i}$.
The ($\phi_{3}$, $r_{B}$, $\delta_{B}$) was extracted from
($x_{-}$, $y_{-}$, $x_{+}$, $y_{+}$) using the frequentist treatment
with the Feldman-Cousins.
The results are
\begin{eqnarray}
\displaystyle
\begin{array}{rcccccccc}
\phi_{3} & = & (77.3  & ^{+}_{-} & ^{15.1}_{14.9} & \pm & 4.2 & \pm & 4.3)^\circ\\
r_{B}    & = & 0.145 & \pm & 0.030 & \pm & 0.011 & \pm & 0.011\\
\delta_{B} & = & (129.9 & \pm & 15.0 & \pm  & 3.9 & \pm & 4.7 )^\circ ,
\end{array}
\end{eqnarray}
where the third error is uncertainty from $c_{i}$ and $s_{i}$.

%%%%%%%%%%%%%%%%%%%%%%%%%%%%%%%%%%%%%%%%%%%%%%%%%%%%%%%%%%%%%%%%%%%%%%%%%
%%
%%   use this format to include a LaTeX table  into your paper
%%
%\begin{table}[!hbtp]
%\begin{center}
%\begin{tabular}{l|cc}  
%\hline\hline
%Event type   &  Number of Events &  Branching Fraction \\ \hline
% Signal 1     &   $2500 \pm 50$  &     $0.25 \pm 0.01$   \\
% Signal 2     &   $100 \pm 10$   &     $0.01 \pm 0.002$  \\
% Background  &    $<1$           &        \\
%\hline\hline
%\end{tabular}
%\caption{Signal and background event yields with statistical errors,
%and branching fraction with statistical and systematic errors added
%in quadrature.}
%\label{tab:results}
%\end{center}
%\end{table}
%%%%%%%%%%%%%%%%%%%%%%%%%%%%%%%%%%%%%%%%%%%%%%%%%%%%%%%%%%%%%%%%%%%%%%%%%%%

\section{Conclusion}

The measurement precision of $\gamma$/$\phi3$ have been progressed 
according to the data accumulation at B-factories, 
and have accelerated by the newly developed efficient physics methods
and analysis techniques. 
Though current statistics of ee colliders is over the 1.2 billion 
$B\bar{B}$ pairs, but it is still too small to exploit the 
$\gamma$/$\phi_{3}$ and the new physics over there.
Current most precise determination is brought by the Dalitz analyses.
Both the model-independent and improved model-dependent analysis pushed down 
the systematic limitation and open up the possibilities of much higher 
precision determination at super B factories in near future.
Furthermore, the recent updated ADS result indicates observation of
the Cabibbo-suppressed $D$ decays which may brought us the competitive
determination with the Dalitz analyses.
Needless to say, it is important the various approaches including GLW 
method should be performed since single analysis can't constrain the 
$\gamma$/$\phi_{3}$ together with the other variables, sufficiently.

\end{document}